\documentclass[12pt]{iopart}   
\usepackage{graphicx}
\usepackage{epsfig}
\usepackage{amssymb}
\begin{document}

\title[Mean field treatment of exclusion processes with random-force
disorder]{Mean field treatment of exclusion processes with random-force disorder}

\author{R\'obert Juh\'asz} 
\address{Research Institute for Solid
State Physics and Optics, H-1525 Budapest, P.O.Box 49, Hungary}
\ead{juhasz@szfki.hu} 

\begin{abstract}
The asymmetric simple exclusion process with random-force disorder is studied
within the mean field approximation. 
The stationary current through a domain with reversed bias is analyzed and 
the results are found to be in accordance with earlier intuitive assumptions.  
On the grounds of these results, a phenomenological random barrier model 
is applied in order to describe
quantitatively the coarsening phenomena. Predictions of the theory are 
compared with numerical results obtained by integrating the mean field
evolution equations. 
\end{abstract}

\maketitle

\newcommand{\bc}{\begin{center}}
\newcommand{\ec}{\end{center}}
\newcommand{\be}{\begin{equation}}
\newcommand{\ee}{\end{equation}}
\newcommand{\beqn}{\begin{eqnarray}}
\newcommand{\eeqn}{\end{eqnarray}}

\section{Introduction}
Transport processes in nature, like intracellular transport which is
realized by active motor proteins \cite{motors} are often modeled by
simple exclusion
processes, in which particles residing on the
sites of a lattice hop stochastically to neighboring sites provided 
the target site is empty \cite{mcdonald,spitzer,liggett,zia,schutzreview}. 
For this paradigmatic model of driven interacting particle systems
many exact results are available \cite{be}. 
As most of the real systems are not ideally translationally invariant, for
instance the filaments on which molecular motors move are heterogeneous, 
a challenging problem is the study of spatially inhomogeneous versions of  
exclusion processes \cite{ramaswamy,stanley,tripathy,goldstein,kolwankar,krug2,hs,jsi,jli06,barma,schadschneider}
for which the bulk of results is obtained by phenomenological methods based on
the statistics of extremes, by mean field approximation and by Monte Carlo simulations.
Most works concern the one-dimensional totally asymmetric process where 
particles can hop only in one direction with site dependent quenched random
rates. In such systems clusters of consecutive bonds with low hop rate act as bottlenecks and the stationary state is segregated, i.e. consists of
macroscopic regions of low and high density \cite{tripathy}. 
When the system is started from a state with homogeneous density it undergoes
a coarsening process in which the typical size of low and high density
segments is growing in time \cite{tripathy,krug2}. 
A similar coarsening phenomenon occurs in the partially asymmetric simple
exclusion process with random-force disorder, where 
the direction of the local bias is random \cite{tripathy,krug2,jsi}. 
In this case, clusters of bonds with
reversed bias compared to the global one limit the current and since their extension is unbounded in an
infinite system, they result in that, parallel with the coarsening of the length
scale, the local currents tend to zero in the long time limit
$t\to\infty$ \cite{jsi}.
In the driven phase of this model, a phenomenological random trap description
was developed which relates the coarsening exponents to the dynamical exponent
of random walk in random environment and the predictions of this theory has
been found to be in agreement with Monte Carlo simulations \cite{jsi}.

In this paper we shall investigate this model within a mean-field 
approximation, 
which, to our knowledge, has not been applied to the disordered partially asymmetric model yet. 
Calculating the steady state current through a segment with a reversed bias we
shall argue that, in the driven phase, extreme value statistics of barrier
heights leads to the same dynamical exponents in the less complex mean field
model as those of the original one. 
As opposed to earlier works applying mean field approximation, here we 
focus on the dynamical behavior rather than the steady state. 
The phenomenological predictions will be checked by numerically integrating
the dynamical mean field equations.  

The rest of the paper is organized as follows.         
In Sec. \ref{model} the model is defined in details. 
In Sec. \ref{phen}, the elements of the phenomenological theory of the steady
state are surveyed. Sec. \ref{barrier} is devoted to the analysis of the current through a single barrier within the mean field approximation with different boundary
conditions. The phenomenological theory of the dynamics is reviewed and
applied to the model in
Sec. \ref{dynamics} and the predictions are compared with numerical results in
Sec. \ref{numerical}. Finally, the results are discussed in
Sec. \ref{discussion} and some calculations for the dynamics of 
the pure model are presented in the Appendix.

\section{The model}
\label{model}
 
The disordered partially asymmetric simple exclusion process is defined as
follows. An infinite one-dimensional lattice is given the sites of which are
either empty or occupied by a particle. On this state space a Markov process
is considered in which particles hop independently to an adjacent site
provided that site is empty. The hop rate from site $i$ to site 
$i+1$ ($i-1$) is denoted by $p_i$ ($q_{i-1}$) and  the pairs of rates
($p_i$, $q_{i+1}$) are i.i.d. positive random variables.
Furthermore, we require that $0<{\rm Prob}(p_i<q_{i-1})<1$. In words, the 
local force $F_i\equiv \ln (p_i/q_i)$ acting on particles can be both positive or
negative with finite probabilities.  

In the mean field approximation, the pair correlations of the occupation number
$n_i=0,1$  are neglected meaning that expected values of products of
occupation numbers $\langle n_in_{i+1}\rangle$ are replaced by $\langle
n_i\rangle\langle n_{i+1}\rangle$.
Then the evolution equation for the local density $\rho_i(t)\equiv \langle
n_i(t)\rangle$ reads as 
\be   
\frac{d\rho_i}{dt}= (p_{i-1}\rho_{i-1}+q_{i}\rho_{i+1})(1-\rho_i)
-[p_i(1-\rho_{i+1})+q_{i-1}(1-\rho_{i-1})]\rho_i.
\label{evolution}
\ee
The current through the $i$th bond can be written as  
\be
J_i(t)=p_i\rho_i(1-\rho_{i+1})-q_i\rho_{i+1}(1-\rho_i).
\label{Jdiscrete}
\ee 
Defining the model on a finite ring of sites $L$ rather than on the integers
it has a steady state where the local currents $J_i$ are all equal. This
stationary current is sample-dependent i.e. depends on the set of random hop
rates $\{p_i,q_i\}$. The typical stationary current in the ensemble of 
samples of size $L$ tends to zero in the limit $L\to\infty$ due to the
occurrence of larger and larger domains with reversed local force that control
the current \cite{tripathy,jsi}. 
In the infinite system, the local densities do not converge in the limit
$t\to\infty$ therefore there exists no stationary state. Nevertheless, when the system 
is started e.g. from a homogeneous state, the local currents $J_i(t)$ which
are non-zero for finite times  
all tend to zero in long time limit \cite{jsi}.       
We shall consider the dynamics of this non-stationary process and are
mainly interested in the dependence of the typical current 
\be
J_{\rm typ}(t)=\exp{\left\{\lim_{L\to\infty}\frac{1}{2L+1}\sum_{i=-L}^L\ln|J_i(t)|\}\right\}}
\label{typical}
\ee
on time.
    
Before analyzing the disordered model, we discuss the evolution of the
typical current in the pure model where $p_i=p$, $q_i=q$ for all $i$. 
As it is shown in the Appendix, the typical deviation of the current from the
stationary one ($J_{\rm \infty}$) decays algebraically with the time. The decay exponent depends
on the symmetries of the model. 
If $p=q$ (symmetric simple exclusion process) the typical current decays as 
\be
J_{\rm typ}(t)\sim t^{-3/4}. 
\ee
If $p\neq q$ (asymmetric simple exclusion process) then, for densities
different from $1/2$, the typical current decays as 
\be
(J-J_{\rm \infty})_{\rm typ}(t)\sim t^{-1/3}, 
\ee
while at half-filling $\rho=1/2$ we have 
\be
(J-J_{\rm \infty})_{\rm typ}(t) \sim t^{-2/3}. 
\ee
These latter results follow essentially from the time-dependence of the
typical deviation of the local density from the stationary value calculated by
Burgers \cite{burgers} but for the sake of self-containedness a short heuristic derivation is given in the
Appendix.   

\section{Phenomenological random barrier theory}
\label{phen}

For exclusion processes with random-force
disorder a phenomenological theory exists by which many steady state and
non-stationary properties are successfully described in accordance with
results of Monte Carlo simulations \cite{tripathy,jsi}.
The basic idea is that the random environment (i.e. the series of jump rates)
contains localized trapping regions or barriers in which the local force is
reversed compared to the global one and such regions therefore can maintain a
very low current. These barriers can be defined quantitatively in terms of the
potential $U_i$ which is defined by 
\be 
\Delta U_i=U_{i+1}-U_i=-F_i=\ln (q_i/p_i). 
\label{potential}
\ee     
An interval from site $a$ to site $b$ is said to be an {\it ascending}
interval if and $U_a<U_i<U_b$ for $a<i<b$.
The ascending interval $[a,b]$ is a {\it barrier} if there does not exist a longer
ascending interval which contains $[a,b]$. 
If the average force is non-zero, i.e.  $\overline{\Delta U}\neq 0$, where the overbar denotes averaging over the
distribution of hop rates, the system is in the driven phase and the size of
the barriers has an exponentially decaying distribution and the number of
barriers in a finite system is proportional to the size of the system. 
Each barrier has a maximal carrying capacity and the smallest one among these
values determines the stationary current of the finite system.  
The key question in this theory is how the maximal carrying 
capacity varies with 
the parameters of the barrier, which depends on the particular model. 
In case of the partially asymmetric simple exclusion process this has been
obtained by the following phenomenological arguments. 
The steady state 
of a homogeneous, open system with reversed bias ($q>p$) where particles enter at site $1$ with rate $\alpha$ and
are removed at site $L$ with rate $\beta$ is exactly known \cite{blythe}.
The density profile contains an anti-shock in the middle of the system, which separates a high density
phase on its left hand side where the density is close to one from a low
density phase on its right hand side where the density is close to zero. 
In case of an inhomogeneous barrier the exact steady state is no longer
available but the profile is qualitatively similar to that of the pure case. 
In the steady state, the anti-shock must be located where the potential (measured from the bottom
of the barrier) is half of the total height of the potential, since the
current of a single particle in the low density phase must be equal to the 
current of a single hole in the high density phase. 
Since the distribution of heights of barriers can be calculated, the
distribution of the current in finite systems is obtained by applying the
statistics of extremes \cite{jsi,jli06}.  

We will apply this phenomenological theory to the mean field model defined
above. First we calculate the mean field current through a random
barrier and shall see that it is determined practically by the height of the
barrier as it has been intuitively assumed for the original stochastic model. 
    
\section{Mean field current over random barriers}
\label{barrier}

\subsection{Open boundaries}

Let us consider an open random barrier with $N$ sites and with entrance and
exit rates $\alpha$ and $\beta$, respectively. 
In the steady state, the local densities in the bulk obey the relations 
\be 
p_i\rho_i(1-\rho_{i+1})-q_i\rho_{i+1}(1-\rho_i)=J, 
\label{rec0}
\ee 
where the current $J$ is to be determined. 
Introducing the variables $y_i=\rho_i/(1-\rho_i)$ and $r_i=p_i/q_i$ 
Eq. (\ref{rec0}) takes the form 
\be
y_{i+1}=r_iy_i-Jq_i^{-1}(1+y_i)(1+y_{i+1}).
\label{rec}
\ee
This is a non-linear recursion equation for the densities.  
Let us choose a site where the potential measured from the left end of the
system is roughly half of the total height of the potential barrier and denote this site by $0$. 
As we shall see later this site is in an anti-shock region where the density
is close to $1/2$. Thus $y_0\approx 1$.      
Denoting the term in Eq (\ref{rec}) which is responsible for non-linearity by 
\be
\omega_i\equiv q_i^{-1}(1+y_i)(1+y_{i+1})
\label{omega}
\ee 
and regarding it as if it was a constant, 
the recursion can be formally carried out
starting from site $0$ to the right, i.e. toward the low density phase till the
rightmost site $L$, yielding: 
\be
y_L=\left(\prod_{j=0}^{L-1}r_j\right)
\left[y_0-J\sum_{j=0}^{L-1}\omega_j\prod_{i=0}^{j}r_i^{-1}\right].
\label{formal}
\ee
The current $J$ is simply related to the density at this site as follows: 
\be
J=\beta\rho_L=\beta\frac{y_L}{1+y_L}=\beta y_L+O(y_L^2). 
\ee
Here, we have used that, as we shall see a posteriori, the density $\rho_i$,
as well as $y_i$ decay exponentially with the site index, and they are thus
very small for large $L$. 
Eliminating $y_L$ from the latter two equations, we obtain the following
formal expression for the current:
\be
J=y_0\frac{\prod_{j=0}^{L-1}r_j}{\beta^{-1}+\Delta_L} + O(y_L^2)=
y_0\frac{e^{-(U_L-U_0)}}{\beta^{-1}+\Delta_L} + O(y_L^2),
\label{Jy}
\ee
with 
\be
\Delta_L=\sum_{j=0}^{L-1}\omega_jr_j^{-1}\prod_{i=j}^{L-1}r_i.
\label{delta}
\ee
The variable $\Delta_L$ is a function of the densities $\rho_i$ but, as we
shall show below, it is bounded by a random variable which is 
finite [$O(1)$] in typical barriers. 
Since the second term in the brackets on the r.h.s. of Eq. (\ref{formal}) is 
negative, the inequality 
\be
y_j<y_0\prod_{i=0}^{j-1}r_i=y_0e^{-U_j}
\label{ineq}
\ee 
obviously holds for all $j>0$. 
Here and in the following, the potential at site $0$ is set to zero, 
i.e. $U_0=0$.
Using these inequalities, we can write 
\beqn
0<\Delta_L\le e^{-U_L}
\sum_{j=0}^{L-1}q_j^{-1}(e^{U_{j+1}}+y_0r_j^{-1})(y_0e^{-U_{j+1}}+1)= \nonumber \\
=y_0e^{-U_L}
\sum_{j=0}^{L-1}q_j^{-1}(1+r_j^{-1}+y_0^{-1}e^{U_{j+1}}+y_0e^{-U_{j}}).
\label{deltabound} 
\eeqn
As can be seen, only those sites give an $O(1)$
contribution to this sum at which the magnitude of the potential is close to $U_L$, i.e. 
either $U_{j}\approx U_L$ or $U_{j}\approx -U_L$. 
The main contribution comes from the sites in the end region at which the
potential is close to $U_L$. Since the random potential is, in general, not necessarily  
monotonic there may be also sites far from the end with $|U_j|\approx U_L$. 
Nevertheless, the barriers have a finite (non-vanishing) average slope 
$\overline{U_L/L}$ in the limit $L\to\infty$, therefore the number of such
sites, as well as the random variable $\Delta_L$ is expected to have an $L$-independent limit distribution.    
If the potential does not turn down to the vicinity of $-U_L$, which is the
typical situation, we can obtain an accurate estimate of $\Delta_L$ as
follows. 
In this case the terms $y_i$ and $y_{i+1}$ appearing in Eq. (\ref{delta})
through $\omega_j$ can be neglected according to
inequality (\ref{ineq}). 
This results in the following expression 
\beqn 
\Delta_L^0=\sum_{j=0}^{L-1}p_j^{-1}\prod_{i=j}^{L-1}r_i= \nonumber \\
=q_{L-1}^{-1}+q_{L-2}^{-1}r_{L-1}+q_{L-3}^{-1}r_{L-1}r_{L-2}+\dots 
+ q_0^{-1}r_{L-1}r_{L-2}\dots r_1, 
\label{db1}
\eeqn
which is thus  an accurate estimate of $\Delta_L$ for large $L$
in the case of barriers for which the potential is well separated from 
$-U_L$, i.e. $-U_L\ll U_i$.
Moreover, this sum starting with the term $j=L-1$ as written above is
rapidly converging if the potential is well separated also from $U_L$ apart
from the region close to the end of the system. This is the case for barriers
with monotonic potential. 
In general samples or for finite $L$, $\Delta_L^0$ is a
lower bound on $\Delta_L$.  

Notice that neglecting the terms $y_i$ and $y_{i+1}$ in
Eq. (\ref{rec}) results in a linear recursion which describes
independent random walkers with density $y_i$ at site $i$. 
As a consequence, the sum $\Delta_L^0$ can be related to
properties of random walks, as follows. 
Let us consider a finite lattice with sites $0,1,\dots,L,L+1$ and the same
series of hop rates as given for the exclusion process except that $p_0$ is
set to zero, furthermore $q_L=0$ and $p_L=\beta$. That means, sites $0$ and
$L+1$ are absorbing. Starting at site $1$, the probability that
the walker is absorbed at site $L+1$ when $t\to\infty$ is called persistence
probability and is given by \cite{ir}: 
\be
p_{\rm pers}(L)=
\left[1+\sum_{i=1}^{L-1}\prod_{j=1}^i\frac{q_{j-1}}{p_j}\right]^{-1}.
\label{pers}
\ee  
This can be recast as 
$p_{\rm pers}(L)=p_0^{-1}e^{-U_L}(\Delta_L^0+\beta^{-1})^{-1}$ which leads to
that, whenever the replacement of $\Delta_L$ by $\Delta_L^0$ is justified, the current is
asymptotically proportional to the persistence probability of the
corresponding random walk: 
\be
J^0=y_0p_0p_{\rm pers}(L).
\ee

The expression of the current in Eq. (\ref{Jy}) is still incomplete 
in the sense 
that it contains the variable $y_0$ at the chosen reference site in the
anti-shock region. This can be, however, easily eliminated as follows.  
Introducing the variables $x_i=y_i^{-1}$, one can write the recursion in
Eq. (\ref{rec0}) for decreasing indeces in the following form: 
\be 
x_{i-1}=r_{i-1}x_i-Jq_{i-1}^{-1}(1+x_i)(1+x_{i-1}), 
\ee
which has the same structure as  Eq. (\ref{rec}) for the forward iteration. 
Performing the recursion from the same initial site as for the forward 
iteration to the entrance site indexed by $-L'$, and using the relation between 
$x_{-L'}$ and the current: 
\be 
J=\alpha(1-\rho_{-L'})=\alpha x_{-L'} + O(x^2_{-L'}),
\ee
we obtain an expression for the current analogous to Eq. (\ref{Jy}): 
\be 
J=x_0\frac{e^{U_{L'}}}{\alpha^{-1}+\Delta_{-L'}} + O(x^2_{-L'}),
\label{Jx}
\ee
where 
\be
\Delta_{-L'}=e^{U_{-L'}}\sum_{j=0}^{-L'+1}q_{j-1}^{-1}(1+x_j)(1+x_{j-1})e^{-U_{j-1}}.
\label{deltax}
\ee
Here, $\Delta_{-L'}$ has the same properties as $\Delta_L$, e.g. the linear
contribution for barriers with $U_i\ll -U_{-L'}$ is given by 
\be
\Delta_{-L'}^0=
q_{-L'}^{-1}+q_{-L'+1}^{-1}r_{-L'}+q_{-L'+2}^{-1}r_{-L'}r_{-L'+1}+\dots 
+ q_{-1}^{-1}r_{-L'}r_{-L'+1}\dots r_{-2}. 
\label{db2}
\ee
This can be again related to a persistence problem 
in a finite system with sites $-L'-1,-L',\dots,-1,0$ 
and with hop 
rates $p_i'=q_i$, $q_i'=p_i$ for $i=-2,-3,\dots,-L'$ and 
$q_{-1}'=p_{-L'-1}'=0$, $p_{-1}'=q_{-1}$, $q_{-L'-1}'=\alpha$.
Now, the walker starts at site $-1$
and the probability that it ends up at site $-L'-1$ 
can be written as 
$\overline{p}_{\rm pers}(L')=(p_{-1})^{-1}e^{U_{-L'}}(\Delta_{-L'}^0+\alpha^{-1})^{-1}$.
This leads to $J^0=x_0p_{-1}\overline{p}_{\rm pers}(L')$.

Obviously, the current in Eq. (\ref{Jx}) must be equal to that obtained by the
forward iteration in Eq. (\ref{Jy}). Multiplying the right
hand sides of the two equations and introducing the
total height of the barrier as $U_N=\prod_{j=-L'}^{L-1}r_i=U_L+|U_{-L'}|$, we
obtain the following
formal expression for the stationary current:
\be
J=\left[(\alpha^{-1}+\Delta_{-L'})(\beta^{-1}+\Delta_{L})\right]^{-1/2}e^{-U_N/2} +
O(J^2).
\label{Jopen}
\ee
Although $\Delta_L$ and $\Delta_{-L'}$ in this expression are given in
terms of the density profile  $\{\rho_i\}$ which is not known exactly
in a closed form, they are bounded by random variables which are 
typically $O(1)$. Moreover, if relations
\beqn 
U_i\gg U_{-L'}  \qquad {\rm for} \quad i>0, \nonumber \\
U_i\ll U_{L}  \qquad {\rm for} \quad i<0 
\label{monotonic}
\eeqn 
are satisfied then $\Delta_L$ and $\Delta_{-L'}$ are accurately approximated
by the linear contributions given in Eqs. (\ref{db1}) and (\ref{db2}) for
large barriers. 
In this case, the current can also be given in terms of persistence
probabilities of random walks: 
\be 
J^0=\sqrt{p_{-1}p_0}\sqrt{p_{\rm pers}(L)\overline{p}_{\rm pers}(L')}.
\ee
In fact, it is easy to see that the only barriers for which the above approximations are invalid 
are those which have a bulk site with $U_i\approx U_L$ and another one
with $U_j\approx -U_{-L'}$, furthermore $i<j$. 
In all other cases the reference site $0$ to which the summations in
$\Delta_L$ and $\Delta_{-L'}$ go, can be shifted such that sites with 
$U_i\approx U_L$ ($U_i\approx -U_{-L'}$) are on the right (left) hand side of
the reference point. 

In case of a homogeneous barrier with $p_i=p$ and $q_i=q$ the condition
in Eq. (\ref{monotonic}) is obviously fulfilled and the current 
for large $N$ is given by\footnote{The exact current of the asymmetric
  simple exclusion process calculated in Ref. \cite{blythe} differs
  from this mean field current by a factor of $r^{1/4}$.}  
\be 
J(N)\simeq\left[\frac{\alpha\beta(q-p)^2}{(\alpha+q-p)(\beta+q-p)}\right]^{1/2}r^{(N-1)/2}.
\ee
For this asymptotically exact mean field current an approximate 
formula has been derived in Ref. \cite{ramaswamy} in the limit of weak asymmetry $p\lesssim q$. 

\subsection{Barrier in an infinite system}

Regarding that the disordered model contains random barriers embedded
in it, we will consider boundary conditions more appropriate for the
above problem, namely the maximal current through a single barrier which is
part of a large disordered system will be analyzed.

Let us assume that the barrier is very far from other barriers,
i.e. the potential is monotonically decreasing outside the barrier. 
Starting the iteration again from a site in the barrier where 
$y_0\approx 1$, we have 
\be 
y_n=e^{-U_n}(y_0-Je^{U_L}\gamma_n),
\label{Jgamma}
\ee
where we have introduced the variables
\be 
\gamma_n = \sum_{j=0}^{n-1}\omega_je^{U_{j+1}-U_L}.
\label{gamma}
\ee
Outside the barrier, the mass flows with a non-vanishing $O(1)$
velocity and taking into account that the stationary current is $O(e^{-U_L})$
this implies that $y_n=O(e^{-U_L})$ for $n\gg L$. 
Thus, far from the barrier $y_ne^{U_n}$ tends to zero and, in the limit 
$n\to\infty$, we obtain from Eq. (\ref{Jgamma}) 
\be 
J\simeq y_0\frac{e^{-U_L}}{\gamma_{\infty}}
\ee
for large $L$. 
A similar backward iteration yields 
$J\simeq x_0\frac{e^{U_{-L'}}}{\gamma_{-\infty}'}$
where $\gamma_{-\infty}'=\sum_{j=0}^{-\infty}q_{j-1}^{-1}(1+x_j)(1+x_{j-1})e^{U_{-L'}-U_{j-1}}$.
Multiplying the two expressions for the current yields finally 
\be 
J\simeq \left(\gamma_{\infty}\gamma_{-\infty}'\right)^{-1/2}e^{-U_N/2}.
\label{Jinfinite}
\ee
The sum $\gamma_{\infty}$ can be decomposed as 
\be
\gamma_{\infty}=\sum_{j=0}^{L-1}\omega_je^{U_{j+1}-U_L}+\sum_{j=L}^{\infty}\omega_je^{U_{j+1}-U_L}=\Delta_L+\sum_{j=L}^{\infty}\omega_je^{U_{j+1}-U_L}.
\ee
In the limit of large barriers ($L\to\infty$) we have 
$\omega_j\simeq q_j^{-1}$ for $j\ge L$ and thus 
\be
\gamma_{\infty}\simeq\Delta_L+p_L^{-1}+p_{L+1}^{-1}r_L^{-1}+p_{L+2}^{-1}r_{L+1}^{-1}r_L^{-1}+\dots
\label{agamma}
\ee
The first term on the r.h.s. is the same one that appears in the current 
of an open
barrier while the sum of the other terms converges since outside the barrier
the potential decreases monotonically ($r_i>1$).  
In case condition (\ref{monotonic}) is met such as for monotonic
barriers, the term $\Delta_L$ in the asymptotic form in Eq. (\ref{agamma}) can
be replaced by $\Delta_L^0$. 
Furthermore, in this case $\gamma_{\infty}$ is related to the persistence
problem in a semi-infinite lattice with sites $0,1,2,\dots$ where the
walker starts at site $1$. The probability that, for $t\to\infty$,
the walker is not at site $0$ can be written as 
$p_{\rm pers}(\infty)\equiv\lim_{L\to\infty}p_{\rm pers}(L)=(p_0e^{U_L}\gamma^0_{\infty})^{-1}$.

The sum $\gamma_{-\infty}'$ can be written in a similar form: 
$\gamma_{-\infty}'\simeq\Delta_{-L'}+p_{-L'-1}^{-1}+p_{-L'-2}^{-1}r_{-L'-1}^{-1}+p_{-L'-3}^{-1}r_{-L'-2}^{-1}r_{-L'-1}^{-1}+\dots$
and for the linear contribution we have the relation
$\overline{p}_{\rm pers}(\infty)\equiv\lim_{L'\to\infty}\overline{p}_{\rm pers}(L')=[p_{-1}e^{-U_{-L'}}(\gamma_{-\infty}')^0]^{-1}$.

Let us now consider a homogeneous barrier where $p_i=p$ and $q_i=q$ if $-L'\le
i<L$ and $p_i=q$ and $q_i=p$ otherwise. 
In this simple case the asymptotic forms of $\gamma_{\infty}$ and $\gamma_{-\infty}'$ can be easily
evaluated yielding the current for large $N$: 
\be
J(N)\simeq\frac{q-p}{2}r^{(N-1)/2}.
\ee

\subsection{A more general model}

As can be seen in the form of the current in Eq. (\ref{Jdiscrete}), the factors
of the form $1-\rho_i$ ensure that the local density at any site, provided it
was initially below $1$, cannot exceed this limit. This is the way how the
mean field approach accounts for the exclusion interaction of the original
process.   
This, however, not the only way to realize hindrance of the flow by the
occupancy of the target site. Remaining at the factorized character of the
mean field current, one could use instead of $1-\rho$ an arbitrary function $\zeta(\rho)$ of the
density of the target site for which the following properties are required. 
First, $\zeta(0)=1$, which means that, if the target site is empty, there is no
hindrance for the current. 
Second, $\zeta(\rho)$ is continuous and monotonically decreasing
with $\rho$ and finally $\zeta(1)=0$, which is responsible for exclusion. 
The dynamics of this general exclusion model is defined by the equations:   
\be   
\frac{d\rho_i}{dt}= (p_{i-1}\rho_{i-1}+q_{i}\rho_{i+1})\zeta(\rho_i)
-[p_i\zeta(\rho_{i+1})+q_{i-1}\zeta(\rho_{i-1})]\rho_i.
\label{gen}
\ee
Our aim with the generalization of the original model is to point out that the
concrete form of $\zeta(\rho)$ is irrelevant regarding the dynamics of the
system in the sense that it influences only the random prefactor in the
expression of the current through a barrier. 

To see this, the calculations of the previous sections can be carried out with
slight modifications. 
With the variable $y_i=\rho_i/\zeta(\rho_i)$ we obtain in the steady state: 
\be
y_{i+1}=r_iy_i-Jq_i^{-1}[\zeta(\rho_i)\zeta(\rho_{i+1})]^{-1}. 
\label{recgen}
\ee
This leads to the same formula as given in Eq. (\ref{Jy}), however, with 
\be
\Delta_L= e^{-U_L}
\sum_{j=0}^{L-1}q_j^{-1}[\zeta(\rho_i)\zeta(\rho_{i+1})]^{-1}e^{U_{j+1}}.
\label{deltagen}
\ee
To obtain an upper bound on $\Delta_L$ we can use inequality (\ref{ineq}) which
still holds. 
First, let us consider the factors $[\zeta(\rho_j)]^{-1}$ where $\rho_j>1/2$.
For these factors we can write: 
$[\zeta(\rho_j)]^{-1}=y_j/\rho_j<2y_j<2y_0e^{-U_j}$. 
For the factors with $\rho_j\le 1/2$ we can use the monotonicity of
$\zeta(\rho)$ to obtain an upper bound: 
$[\zeta(\rho_j)]^{-1}\le [\zeta(1/2)]^{-1}=const$.
We have thus $[\zeta(\rho_j)]^{-1}\le \max\{2y_0e^{-U_j},[\zeta(1/2)]^{-1}\}$.
Using these inequalities an upper bound on $\Delta_L$ is obtained which 
contains an $O(1)$ contribution in the sum for sites where the magnitude of the
potential is very close to $U_L$. 
The backward iteration can be done in an analogous way and similar 
conclusions for $\Delta_{-L'}$ can be drawn. 
The final conclusion is that the current can be written in the form given in
Eq. (\ref{Jopen}) in case of an open barrier an in the form given in
Eq. (\ref{Jinfinite}) in case of an infinite system and in both cases the
concrete form of $\zeta(\rho)$ influences only the prefactors in front of the
exponentials.  
Moreover, if relations in Eq. (\ref{monotonic}) are fulfilled, the linear
contributions $\Delta^0_L$ and $\Delta^0_{-L'}$ are
 independent of the form of $\zeta(\rho)$ and are thus given by the expressions
 obtained in the previous section.

\section{Phenomenological theory of the dynamics}
\label{dynamics}

The description of the non-stationary state of the system is based
on that segments of characteristic length $\xi$ can be regarded as 
quasi-stationary and as time elapses the characteristic
length scale $\xi$ increases.  
Thus, the steady state properties of a finite system has to be    
reviewed first. 

\subsection{Driven phase}

Let us assume that the system is driven to the right on
average, i.e. $\overline{\Delta U}<0$.  
In a finite but large system of size $N$, there are $O(N)$
barriers and the stationary current $J(N)$ is roughly 
equal to the smallest one among the currents of barriers $J_i$
considered in the previous section. 
We have obtained there that the current through a barrier 
is $J_i=C_ie^{-U_i/2}$ where $U_i$ is the height of the potential barrier
and $C_i$ is an $O(1)$ random factor which depends on the shape of the
barrier (and that of the environment in close vicinity of the barrier). Although we have assumed
there that the barrier is well separated from other barriers, which
does not hold in a disordered system, the neighboring
barriers are expected to influence only the random prefactor $C_i$. 
The relevant factor in $J_i$ is $e^{-U_i/2}$, the inverse of which is 
roughly the square root
of the waiting time $\tau_i\sim e^{U_i}$ of a single random walker 
at that barrier. The distribution of the random variable $\tau_i$ is known to
have an algebraic tail \cite{bouchaud}
\be 
 P_>(\tau)\simeq A\tau^{-\mu}, 
\label{taudist}
\ee
where the control parameter $\mu$ is the positive root of the equation
\be 
\overline{e^{\mu\Delta U}}=1. 
\label{mu}
\ee
It follows then that the current through barriers $J_i\sim \tau_i^{-1/2}$
has the asymptotic distribution 
\be
P_<(J)\simeq A'J^{2\mu}
\label{Jdist}
\ee
for $J\to 0$. 
The current in a finite system then follows a Fr\'echet distribution
and has the typical value vanishing with $N$ as \cite{jsi,jli06} 
\be
J_{\rm typ}(N)\sim N^{-1/2\mu}. 
\label{JN}
\ee
In the steady state, a phase separation can be observed: almost all
mass accumulates behind the highest barrier and forms a high density
phase of macroscopic size, where the density is close to one; in the rest of
the system the density is close to zero. Within these phases, the density
profile is not completely flat but it contains peaks at those barriers
whose height is greater than the half of the highest potential
barrier. The number of these peaks is $O(N^{1/2})$ where the exponent
$1/2$ is universal in the driven phase and is related to the
half-filling of the highest barrier \cite{jsi}.   

Let us assume now that the system is started from a state with random
local densities and with an average density $1/2$. 
After time $t$ has elapsed, the characteristic length scale is $\xi(t)$ and  
the typical size of high and low density segments is $\xi(t)/2$. 
The rate of growth of these domains is proportional to the typical
current $J_{\rm typ}(t)$ at that time scale, that means we can write
\be 
\frac{d\xi}{dt}\sim J_{\rm typ}[\xi(t)]\sim \xi^{-1/2\mu}. 
\label{coars}
\ee    
Integrating this differential equation yields 
\be 
\xi(t)\sim t^{2\mu/(1+2\mu)}
\ee
and 
\be 
J_{\rm typ}(t)\sim t^{-1/(1+2\mu)}.
\label{Jt}
\ee
So, the typical current decays algebraically just as in the pure model
but with a non-universal decay exponent $\beta=1/(1+2\mu)$. 
The growth of the length scale $\xi(t)$ follows also a power-law. 
It is, however, more convenient to measure the average distance  $l(t)$ between
adjacent peaks of the density profile in
numerical simulations rather than $\xi(t)$. This quantity grows as 
\be 
l(t)\sim\sqrt{\xi(t)}\sim t^{\mu/(1+2\mu)},   
\label{lt}
\ee
again with a non-universal coarsening exponent $\delta=\mu/(1+2\mu)$. 

So far we have tacitly assumed that the barriers comprise many sites such that it is
reasonable to speak of half-filling of barriers.  
This is, however, not always true when the distribution of forward hop rates $p$ is
not bounded away from zero. In that case the barriers may typically consist of
 single links through which the rate of forward hopping is vary small, and as
 a consequence, the above calculations have to be modified. 
Let us assume that the distribution of $p$ has the asymptotic form 
$P_<(p)\simeq const\cdot p^{\nu}$  for $p\to 0$. 
Comparing this to Eq. (\ref{Jdist}), it is clear that whenever $\nu<2\mu$ the
local currents are controlled almost always by barriers consisting of single
links and Eq. (\ref{JN}) changes to
\be
J_{\rm typ}(N)\sim N^{-1/\nu}    \qquad    (\nu<2\mu).
\label{JNa}
\ee  
This anomalous scaling of the stationary current has been revealed in Ref. \cite{jsi}.
Here we go further and derive how the dynamics are modified if
$\nu<2\mu$. 
Using Eq. (\ref{JNa}), the evolution equation 
$\dot{\xi}\sim J_{\rm typ}[\xi(t)]$
for the typical length of quasistationary segments results in 
\be
\xi(t)\sim t^{\nu/(1+\nu)}.
\ee
This relation together with Eq. (\ref{JNa}) yields the following time
dependence of the typical current: 
\be 
J_{\rm typ}(t)\sim t^{-1/(1+\nu)}    \qquad    (\nu<2\mu).
\label{Jta}
\ee 
In a segment of size $\xi$, where the quasistationary current is 
$J(\xi)\sim\xi^{-1/\nu}$, 
mass accumulates at those extended barriers where the waiting time 
$\tau_i\sim e^{U_i}$ is greater than $1/J(\xi)$. 
Making use of the distribution of waiting times in Eq. (\ref{taudist}) 
we obtain that the number $n(\xi)$ of such
barriers in the segment scales as 
\be 
n(\xi)\sim \xi P_>[1/J(\xi)]\sim \xi^{1-\mu/\nu}.
\ee
Thus, the length scale $l(t)$ grows with time as 
\be 
l(t)\sim \xi/n(\xi)\sim \xi^{\mu/\nu}\sim t^{\mu/(1+\nu)}
\qquad  (\nu<2\mu).
\label{lta}
\ee

\subsection{Zero average force}

If the average force is zero, i.e. 
$\overline{\Delta U}=0$, then the extension of the largest barrier is $O(N)$
and the above theory breaks down.  
At this point we have only scaling considerations at our disposal \cite{jsi}. 
The height of the largest barrier is $O(\sqrt{N})$, therefore the typical
stationary current in a finite system is expected to scale with $N$ as  
\be 
-\ln J_{\rm typ}(N)\sim \sqrt{N}.
\ee
Plugging this relation into the r.h.s. of Eq. (\ref{coars}) yields 
\be 
\xi(t)\sim\left[\ln\left(t/\ln t\right)\right]^2
\label{xit}
\ee
and 
\be
J_{\rm typ}(t)\sim t^{-1}\ln t.
\label{Jtc}
\ee
In the steady state of a finite system almost all mass is concentrated in $O(\sqrt{N})$ basins
where the density is close to one and the extension of the largest basin is
$O(N)$ \cite{jsi}.  
Thus the number of jumps in the density profile, where the density crosses
over from high ($\rho\approx 1$) to low ($\rho\approx 0$) density is 
$O(\sqrt{N})$. 
Defining $l(t)$ as the average distance between adjacent jumps in the profile
at time $t$ and assuming quasi-stationarity in segments of characteristic size
$\xi(t)$, we obtain that it increases with time as 
\be 
l(t)\sim\sqrt{\xi(t)}\sim\ln\left(t/\ln t\right).
\label{ltc}
\ee

In case of zero average force, we have formally $\mu=0$ from
Eq. (\ref{mu}). The formulae for the dynamical quantities obtained here are
consistent (apart from logarithmic factors) with those valid in the driven
phase taken in the limit $\mu\to 0$. 

\section{Numerical analysis}
\label{numerical}
 
The stationary properties of the disordered model predicted by the
phenomenological theory has been compared with
results of Monte Carlo simulations and a good agreement has been found
\cite{jsi}. 
The dynamical behavior of the current and the length scale has not been
directly checked. The reason for this is
that even for each random sample many runs have to be performed with different
stochastic histories in order to calculate to local currents or the density
profile. 
Instead of this, the time dependence of the displacement of a tagged particle has been
measured in case of zero average force \cite{stanley} and in the driven phase \cite{jsi}.  
By solving the evolution equations (\ref{evolution}) in the mean 
field treatment, the local densities and currents are directly at our
disposal and the dynamical behavior of $J_{\rm typ}(t)$ and $l(t)$ can be
conveniently checked. 

In the numerical calculations, we have considered two types of distributions for
the hop rates.
A discrete one, where $p_i+q_ {i+1}=1$ and the probability density of $p$ is 
\be
f(p)=c\delta[r/(1+r)-p]+(1-c)\delta[1/(1+r)-p],
\label{binary}
\ee
where $0<c\le 1/2$ and  $0<r<1$ are constants, 
and a continuous one with probability densities
\beqn 
f(p)=1/s\quad {\rm if}\quad 0\le p\le s\quad {\rm and}\quad f(p)=0\quad{\rm
  otherwise;} \nonumber \\
g(q)=1\quad {\rm if}\quad 0\le q\le 1\quad {\rm and}\quad g(q)=0\quad{\rm
  otherwise.}
\label{continuous}
\eeqn
The control parameter $\mu$ is given by 
\be 
\mu=\frac{\ln(c^{-1}-1)}{\ln(1/r)}
\label{mu_bin}
\ee
in the former case and implicitly by 
\be 
s=(1-\mu^2)^{-1/\mu}
\label{mu_cont}
\ee
in the latter case. 
In the case of the continuous randomness, the anomalous scaling given in Eqs. 
(\ref{Jta}) and (\ref{lta}) sets in with $\nu=1$ if $\mu>1/2$, while for the
discrete randomness the scaling is never anomalous.

We have generated random samples of size $L=10^5-10^6$ and starting from a
disordered initial state where the local densities are independent random
variables with a homogeneous distribution in the range $[0,1]$,  
the evolution equations in Eq. (\ref{evolution}) have been numerically integrated by the $4$th order
Runge-Kutta method \cite{numrec} up to time $2^{19}$.
For the times where measurements were carried out, 
the coarsening length scale was much less than the size of
the system so that the system can be practically regarded as infinite.  
We have calculated the time-dependence of the finite-$L$ estimate of the
typical current given in Eq. (\ref{typical}) and the time-dependence of the 
length $l(t)=L/n(t)$, where $n(t)$ is the number of points where the density
profile crosses the line $\rho=1/2$. 
These calculations have been repeated for $10^2$ independent random samples and
the averages of the above quantities have been calculated. 
Having the measured data $J_{\rm typ}(t_n)$ and 
$l(t_n)$, we have calculated effective exponents from
neighboring data points at time $t_n$ and $t_{n+1}$: 
\be
\beta_{\rm eff}(t_n)=\frac{\ln[J_{\rm typ}(t_{n+1})/J_{\rm typ}(t_n)]}
{\ln[t_{n+1}/t_n]},
\qquad 
\delta_{\rm eff}(t_n)=\frac{\ln[l(t_{n+1})/l(t_n)]}
{\ln[t_{n+1}/t_n]}.
\label{effexp}
\ee
In addition to this, we have also investigated the distribution of local currents. 

\begin{figure}[h]
\includegraphics[width=0.5\linewidth]{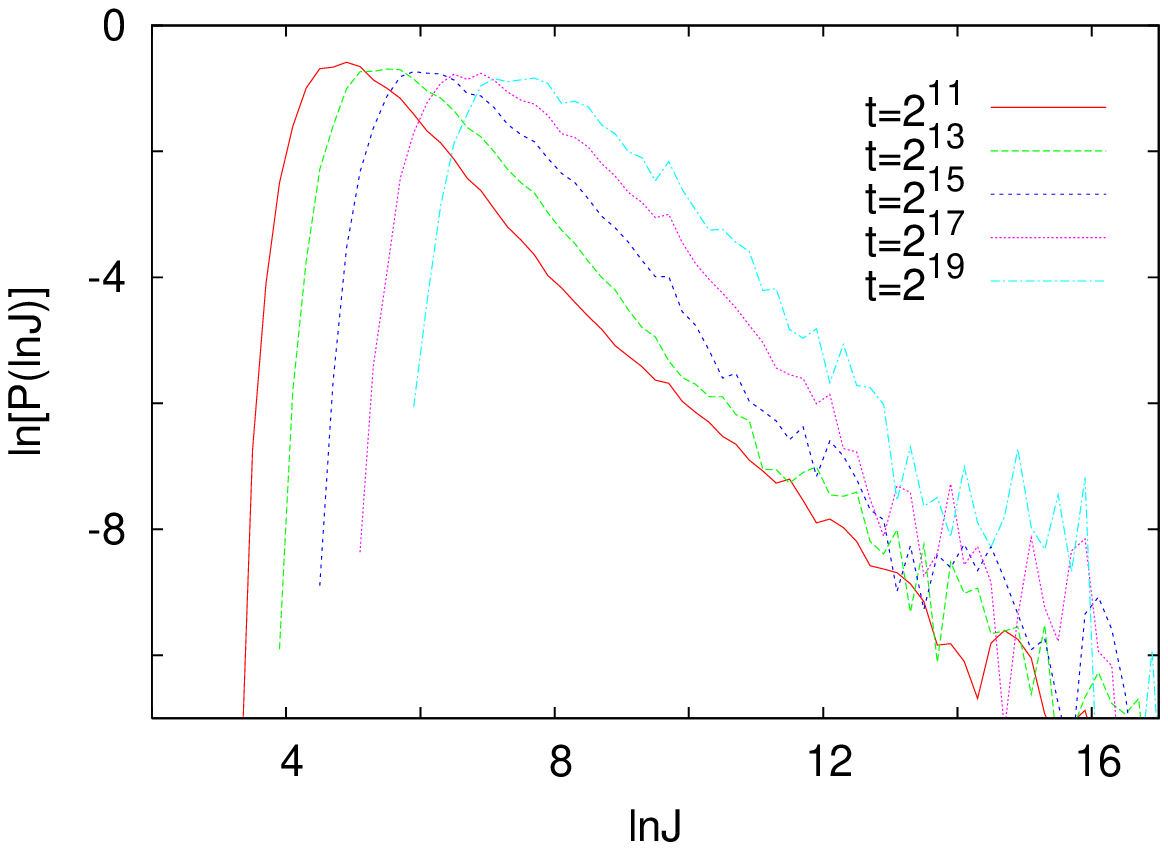}
\includegraphics[width=0.5\linewidth]{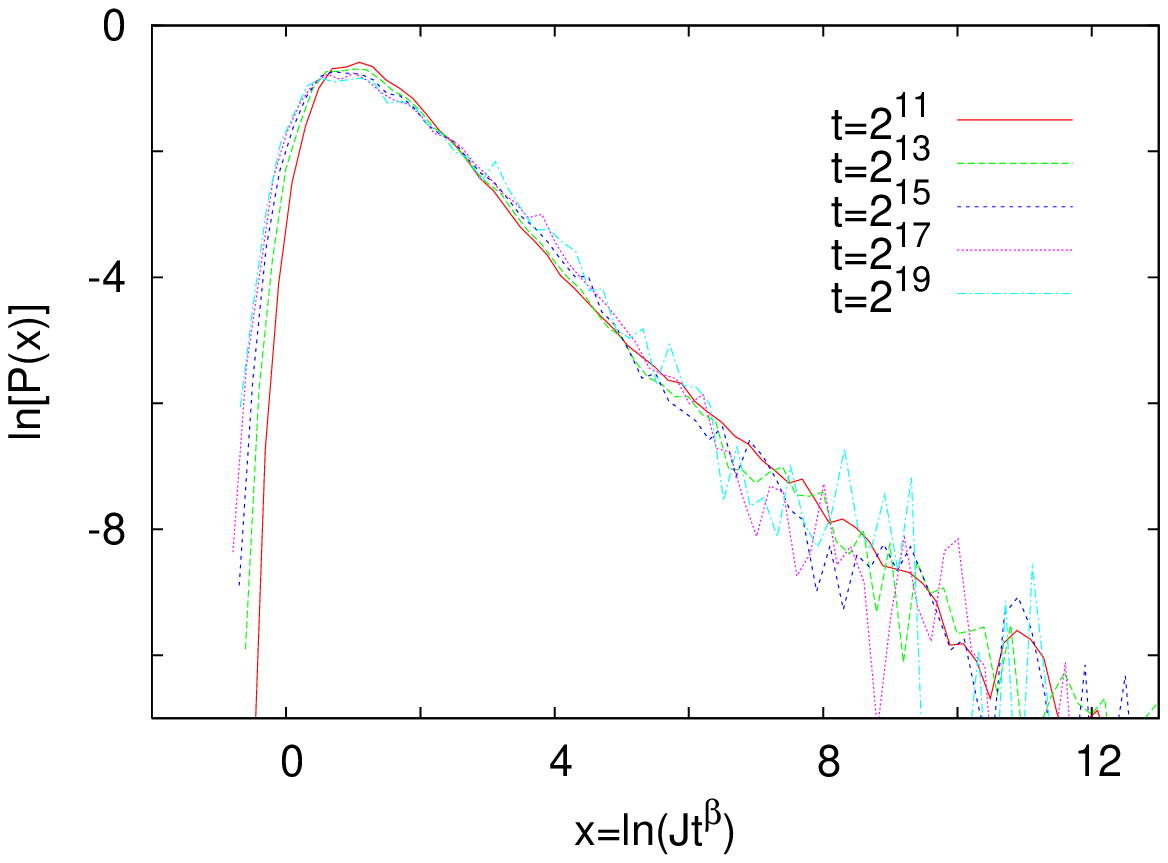}
\caption{\label{fig1} Left: Numerically calculated distribution of the
  logarithm of local currents at different times for the binary randomness with
  parameters $r=1/4$, $c=1/3$. The control parameter is $\mu=1/2$ and
  $\beta=1/2$. Right: Scaling plot of the same data.}
\end{figure}

\begin{figure}[h]
\includegraphics[width=0.5\linewidth]{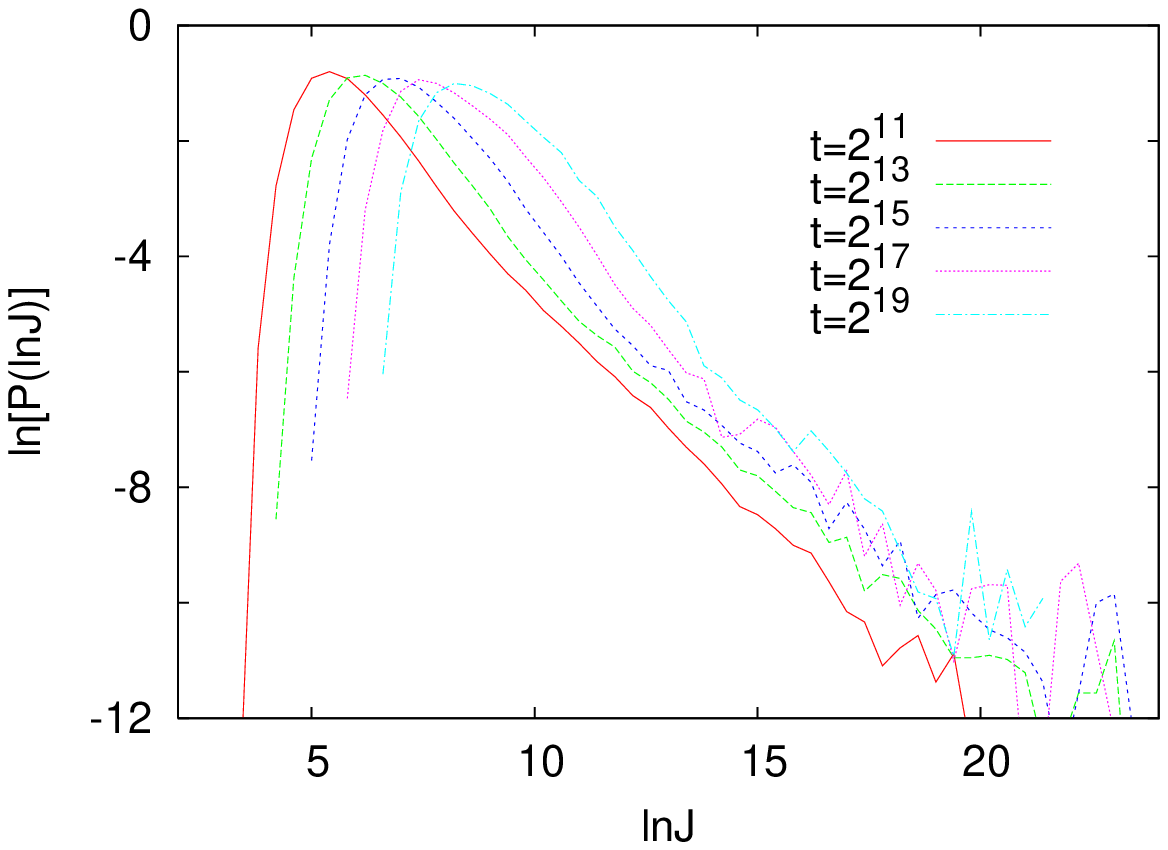}
\includegraphics[width=0.5\linewidth]{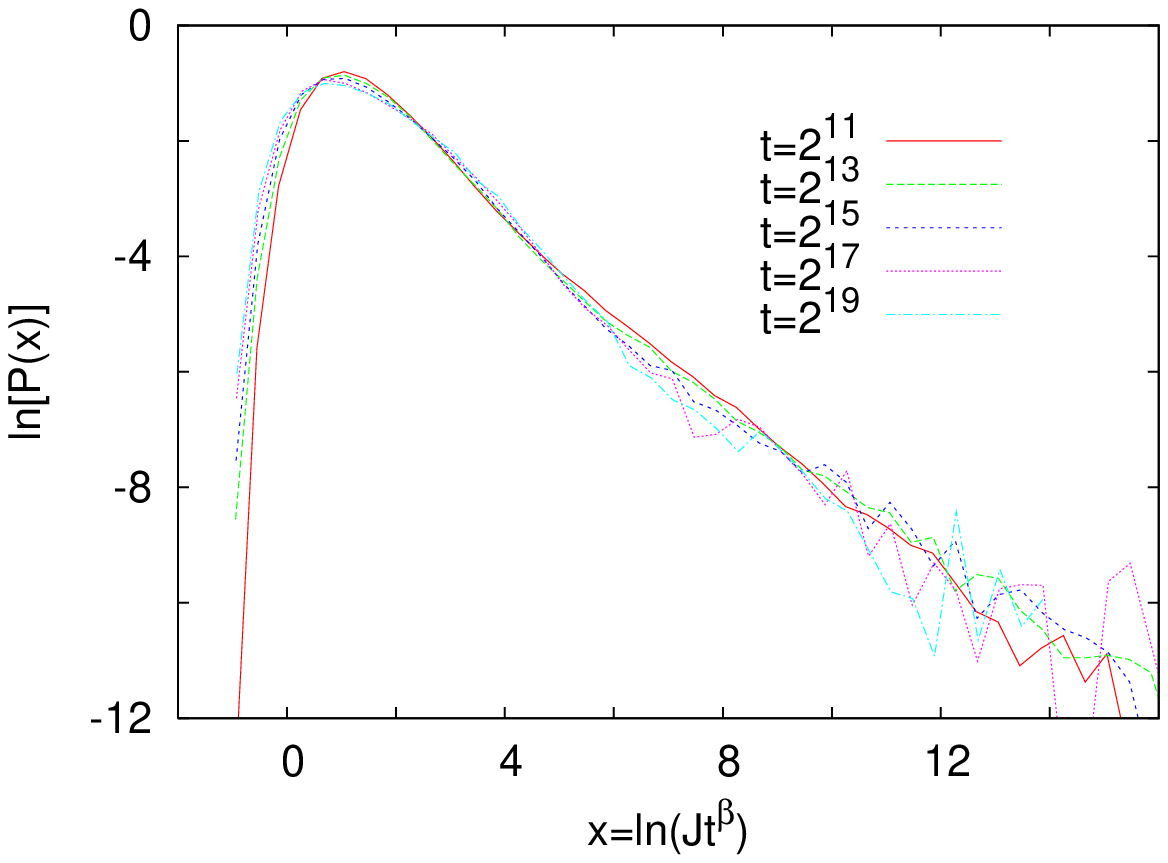}
\caption{\label{fig2} Left: Numerically calculated distribution of the
  logarithm of local currents at different times for 
the uniform randomness with  $s=1.5$. 
The control parameter is $\mu\approx 0.37607$ and $\beta\approx 0.570734$. 
Right: Scaling plot of the same data.}
\end{figure}

We start the presentation of numerical results with the driven phase, where
$\mu>0$. 
The distribution of local currents at different times can be seen in Fig. \ref{fig1} for the
binary randomness and in  Fig. \ref{fig2} for the uniform one. 

\begin{figure}[h]
\includegraphics[width=0.5\linewidth]{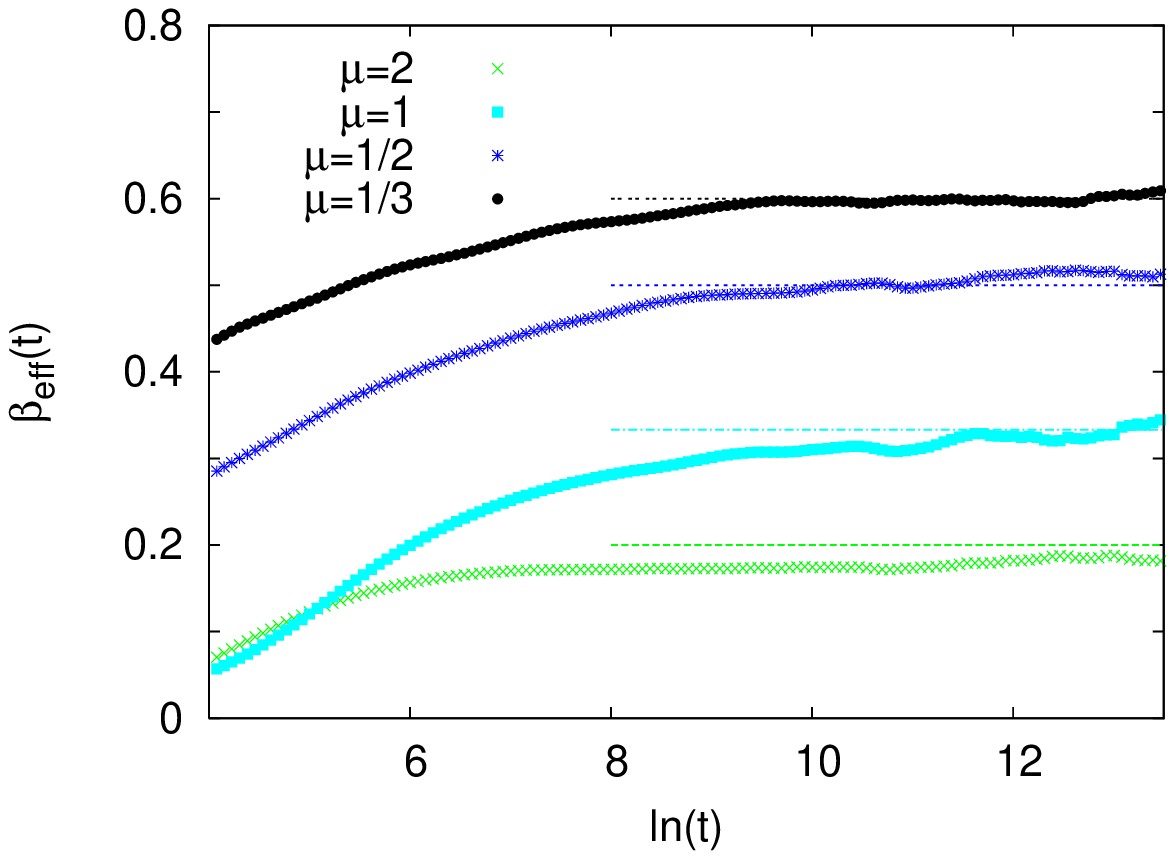}
\includegraphics[width=0.5\linewidth]{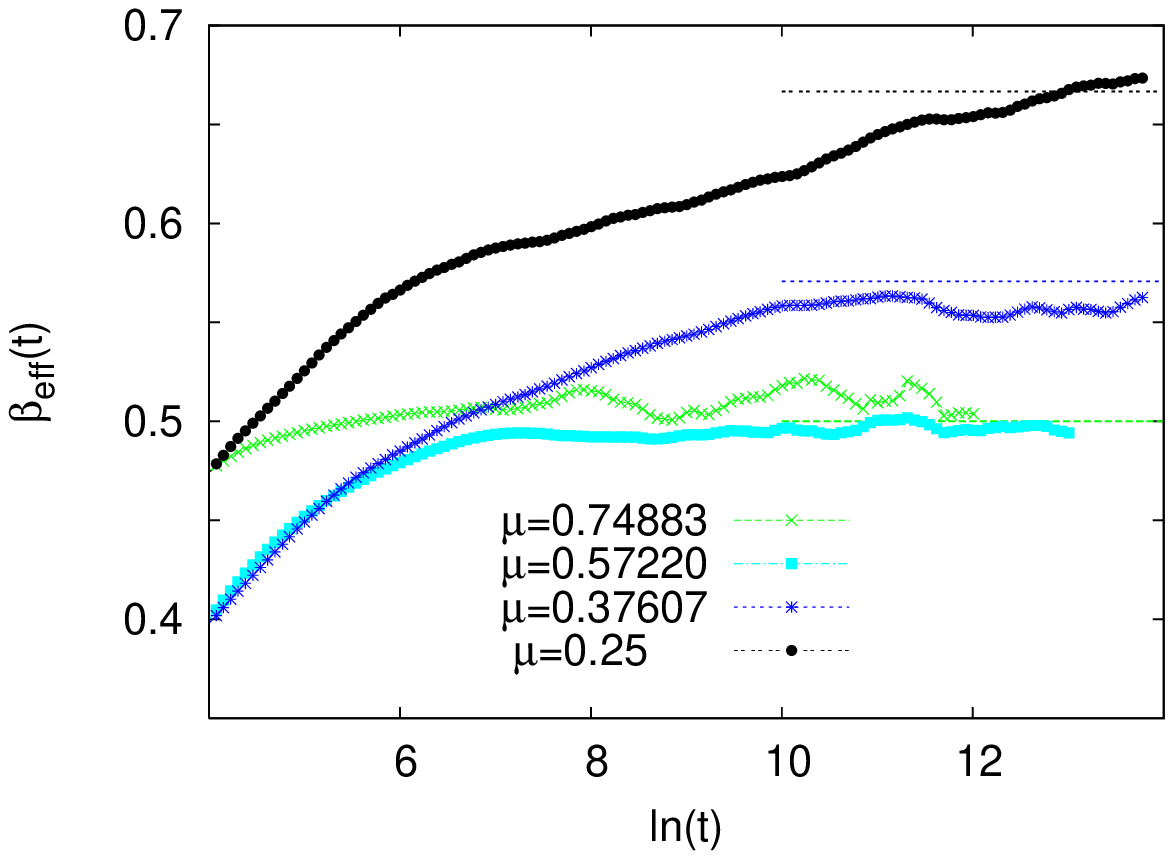}
\caption{\label{fig3} Effective exponent $\beta_{\rm eff}(t)$ plotted
  against time in different points of the driven phase. Left: Data are
  obtained with the binary randomness with parameters $c=0.2$, $r=0.5$ and
  $c=1/3$, $r=0.5,0.25,0.125$ where the control parameter is 
$\mu=2,1,1/2,1/3$, respectively. 
Right: Data are obtained with the uniform randomness with $s=3,2,1.5,1.29$
where $\mu\approx 0.7488,0.5722,0.3761,0.25$, respectively. 
The horizontal lines indicate the value of $\beta$ predicted the phenomenological theory.}
\end{figure}

As can be seen in the figures, an adequate data collapsing can be 
achieved using the scaling variable $Jt^{\beta}$ where $\beta$ is the exponent
predicted by the theory in Eq. (\ref{Jt}).   
The time-dependence of the typical current has been calculated in several
points of the driven phase and the corresponding effective exponents
$\beta_{\rm eff}(t)$ are plotted against time in Fig. \ref{fig3}.
The obtained data are again in satisfactory agreement with the predictions of phenomenological theory. 

\begin{figure}[h]
\includegraphics[width=0.5\linewidth]{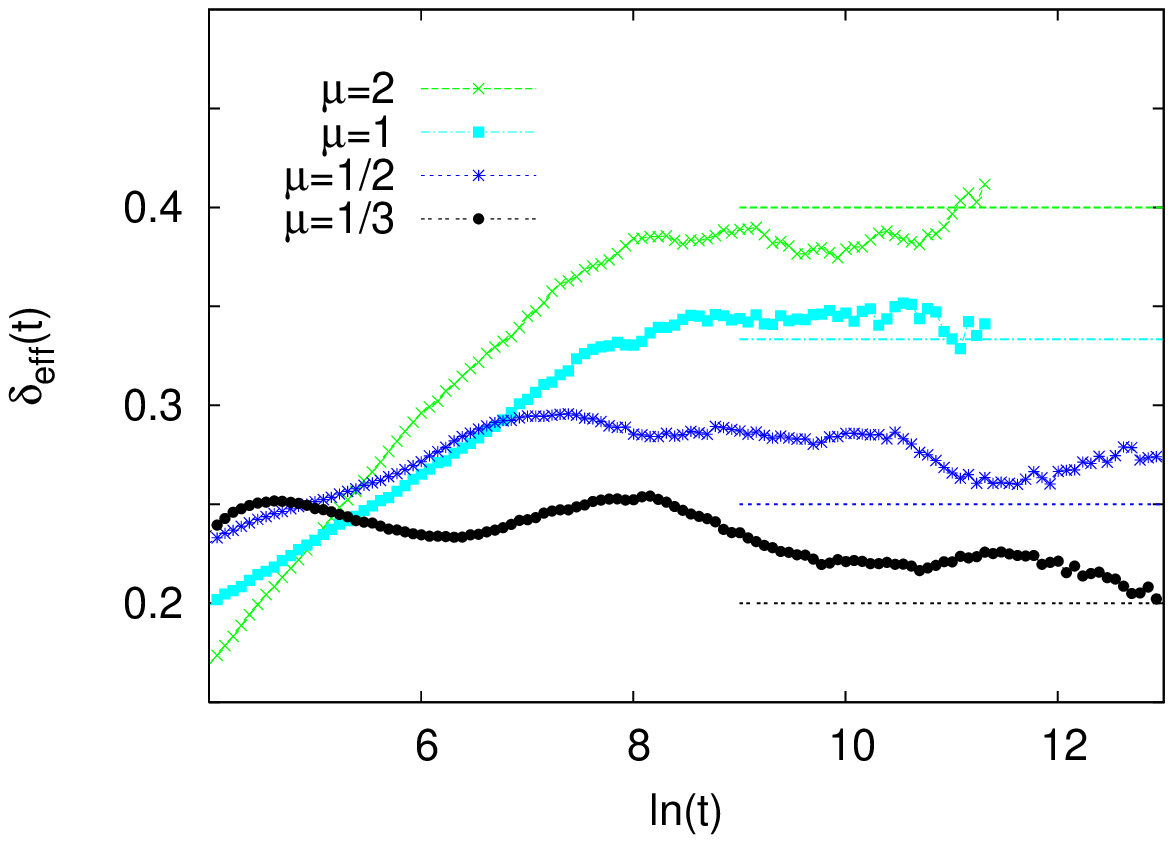}
\includegraphics[width=0.5\linewidth]{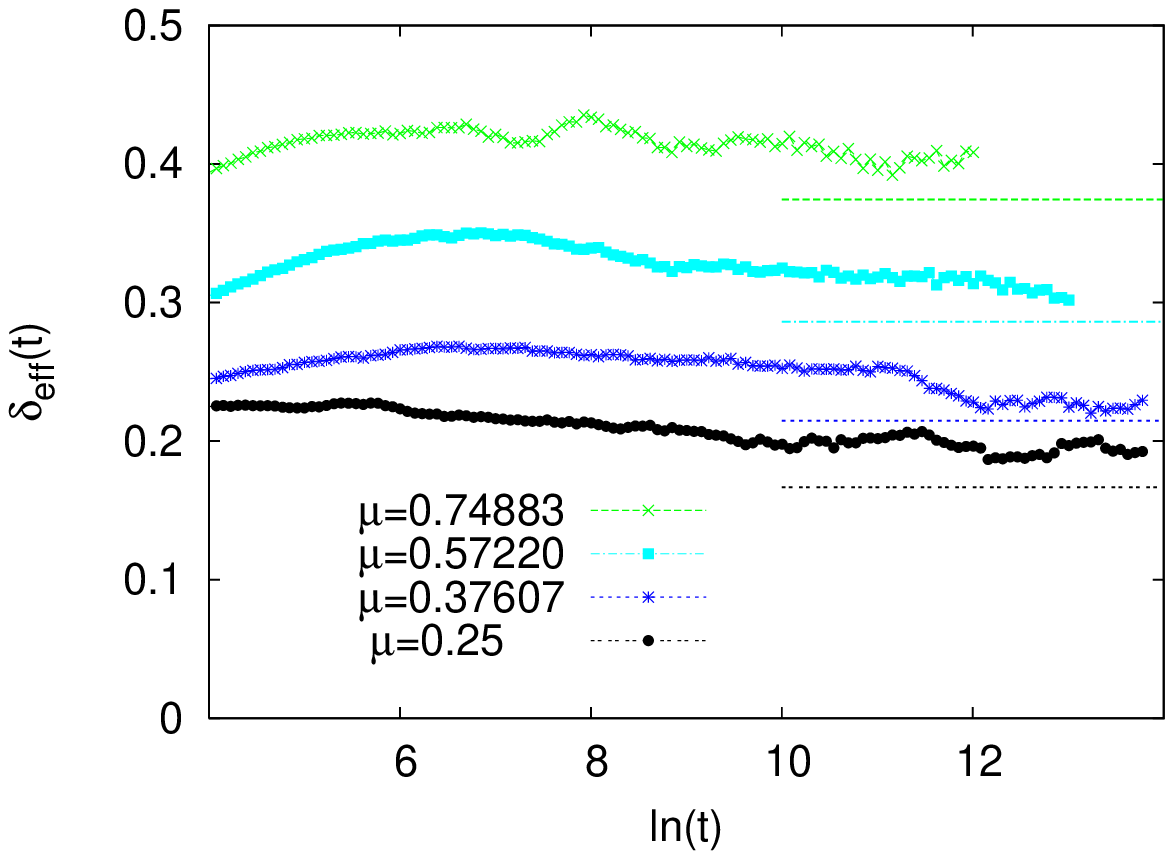}
\caption{\label{fig4}
Effective exponent $\delta_{\rm eff}(t)$ plotted
  against time in different points of the driven phase. Left: Data are
  obtained with binary randomness with parameters $c=0.2$, $r=0.5$ and
  $c=1/3$, $r=0.5,0.25,0.125$ where the control parameter is 
$\mu=2,1,1/2,1/3$, respectively. 
Right: Data are obtained with the uniform randomness with $s=3,2,1.5,1.29$
where $\mu\approx 0.7488,0.5722,0.3761,0.25$, respectively. 
The horizontal lines indicate the value of $\delta$ predicted the phenomenological theory.
}
\end{figure}

The length $l(t)$ has been measured at the same points of the driven phase, as
well. 
The corresponding effective exponents are compared to the predictions of
phenomenological theory in Fig. \ref{fig4}. As can be seen, the finite time
corrections are more considerable than those of $\beta$, nevertheless
the asymptotic behavior is still compatible with the theory. 

\begin{figure}[h]
\includegraphics[width=0.5\linewidth]{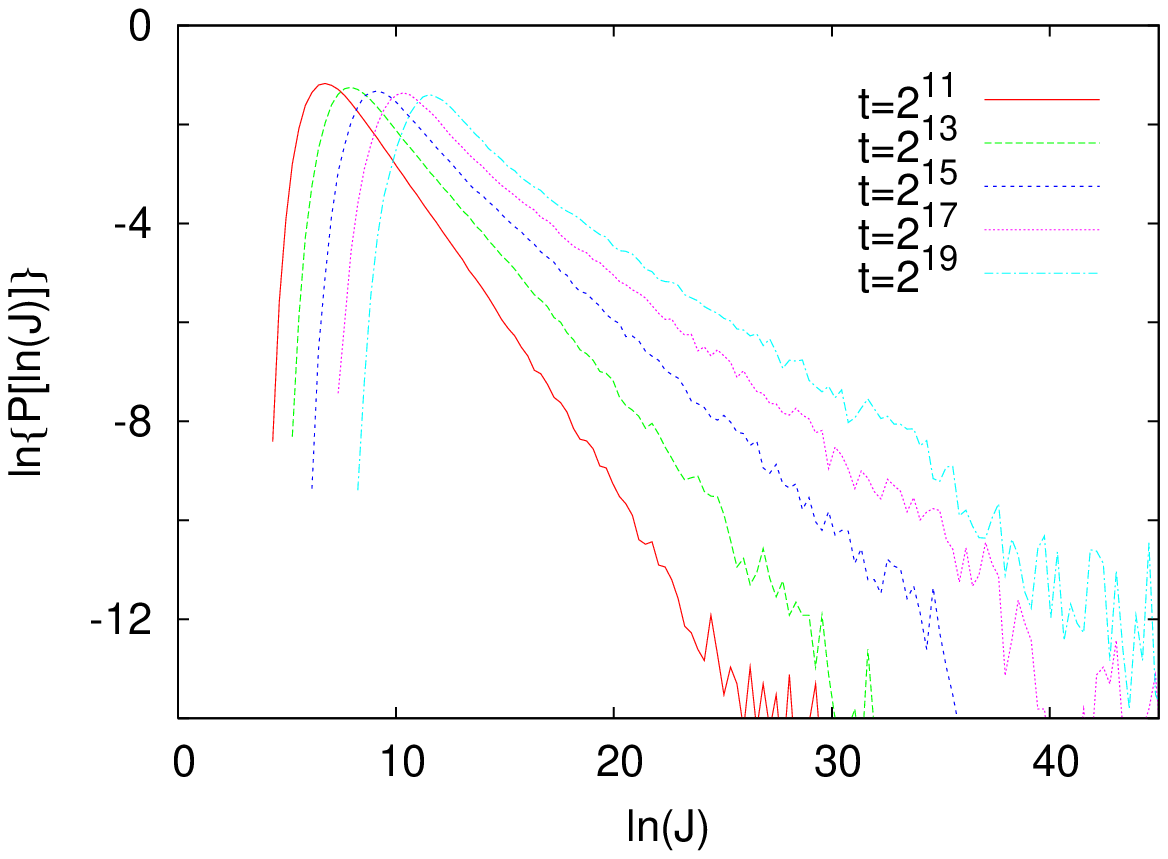}
\includegraphics[width=0.5\linewidth]{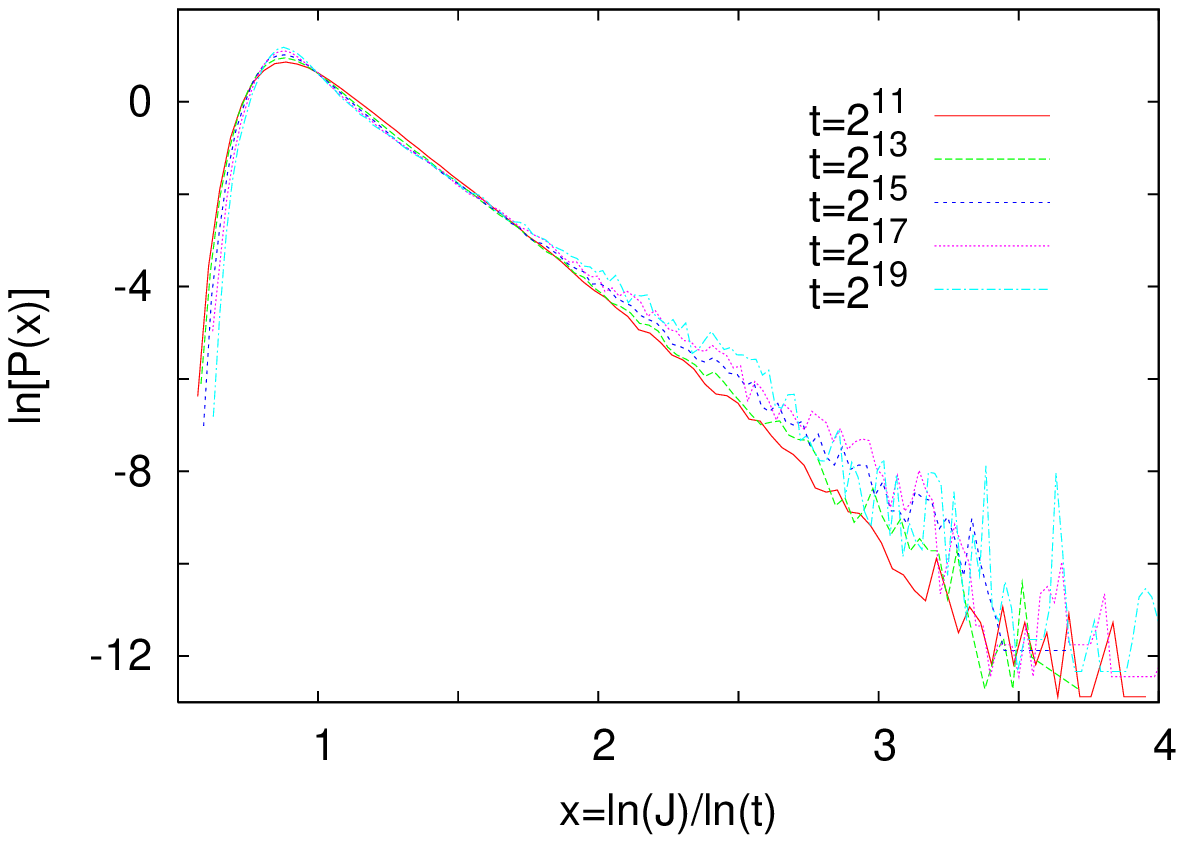}
\caption{\label{fig5} Left: Numerically calculated distribution of the
  logarithm of local currents at different times for the binary randomness with
  parameters $r=1/4$, $c=1/2$. At this point the average force is zero 
and $\mu=0$. Right: Scaling plot of the same data.
}
\end{figure}

\begin{figure}[h]
\includegraphics[width=0.5\linewidth]{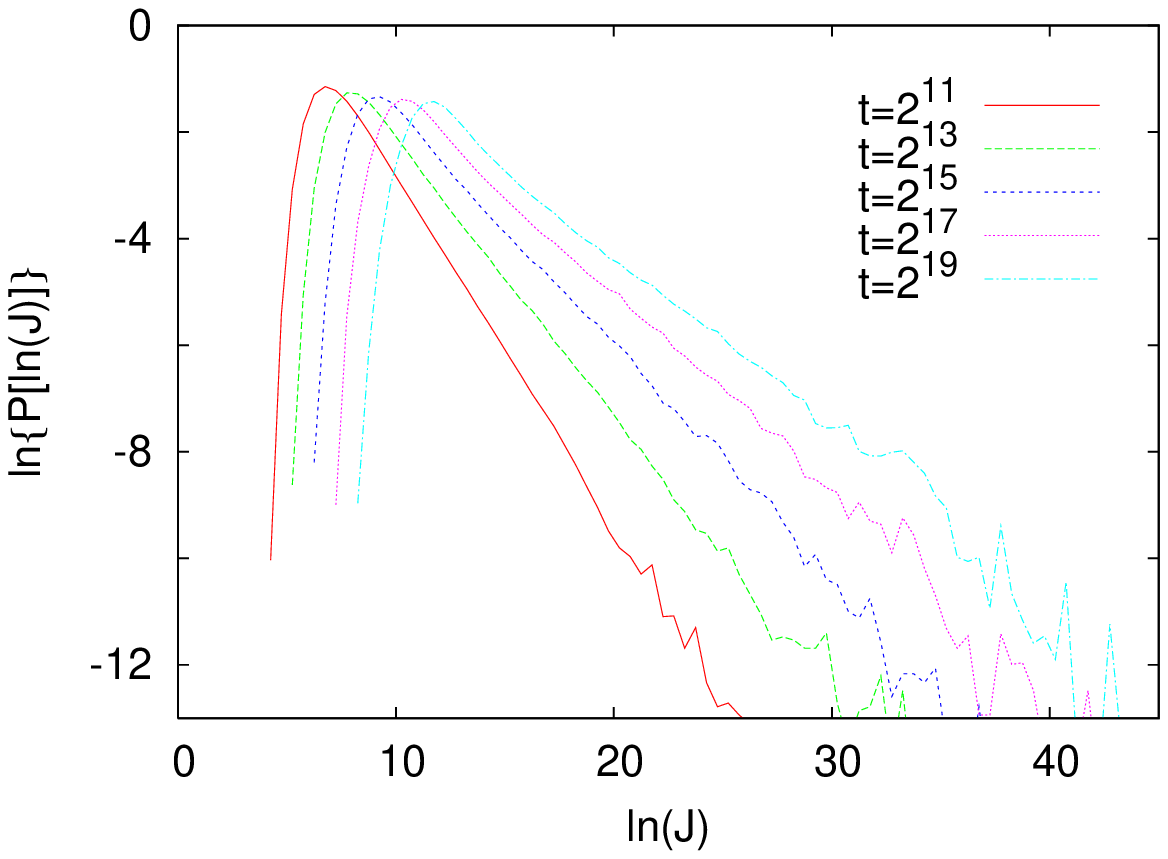}
\includegraphics[width=0.5\linewidth]{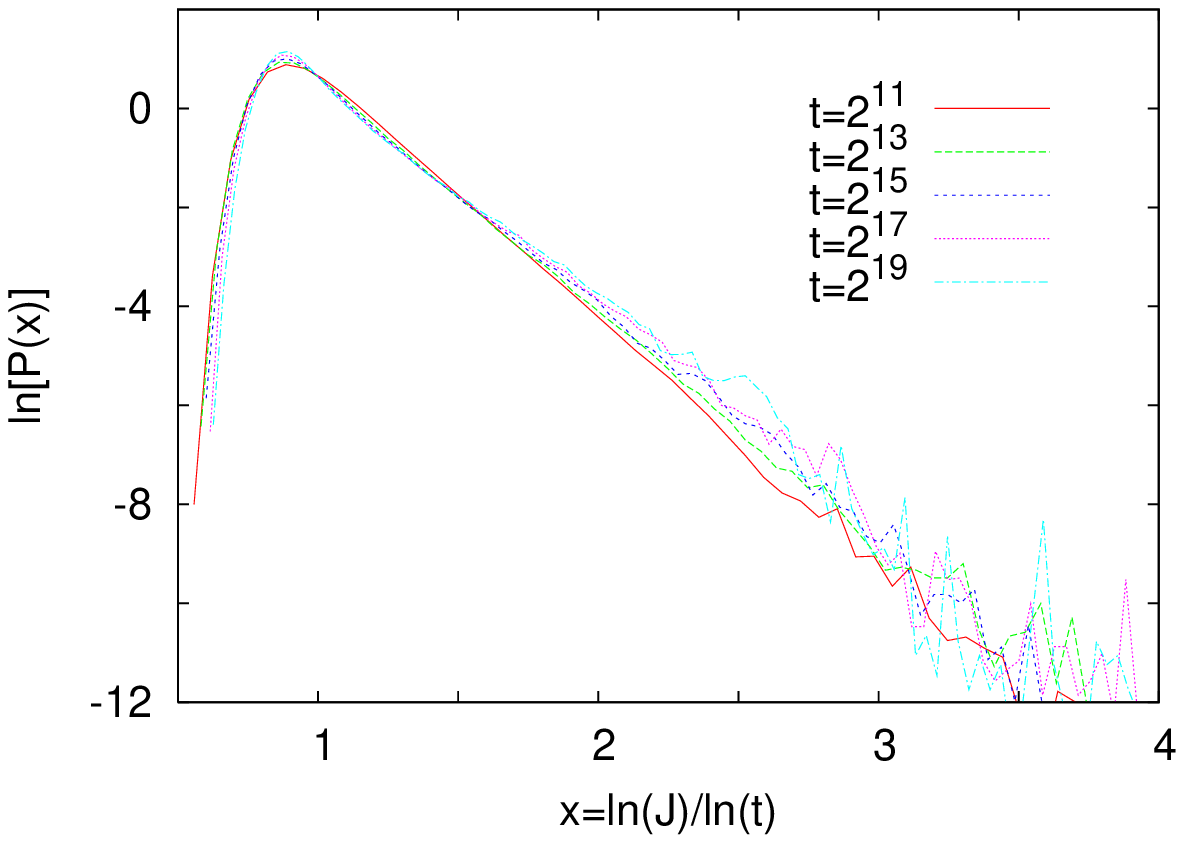}
\caption{\label{fig6} Left: Numerically calculated distribution of the
  logarithm of local currents at different times for uniform randomness with
  parameters $s=1$. At this point the average force is zero 
and $\mu=0$. Right: Scaling plot of the same data.}
\end{figure}

\begin{figure}[h]
\includegraphics[width=0.5\linewidth]{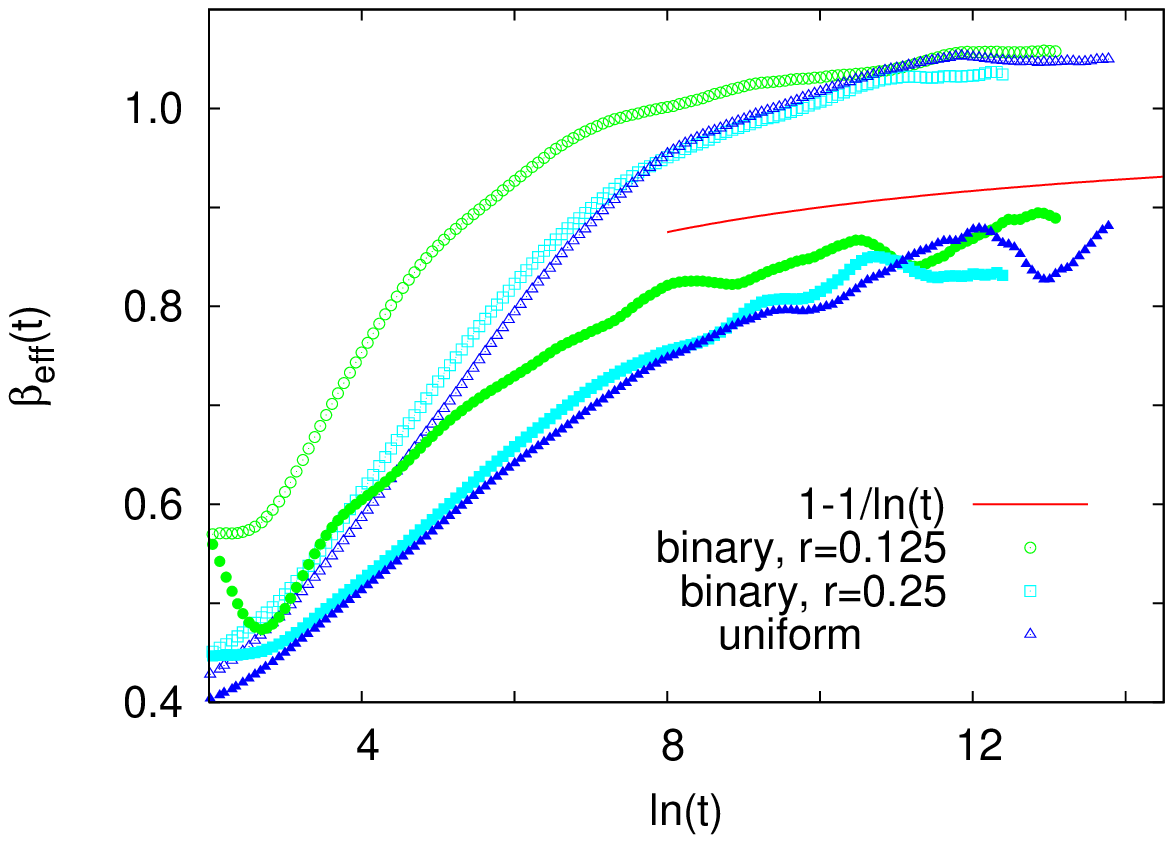}
\includegraphics[width=0.5\linewidth]{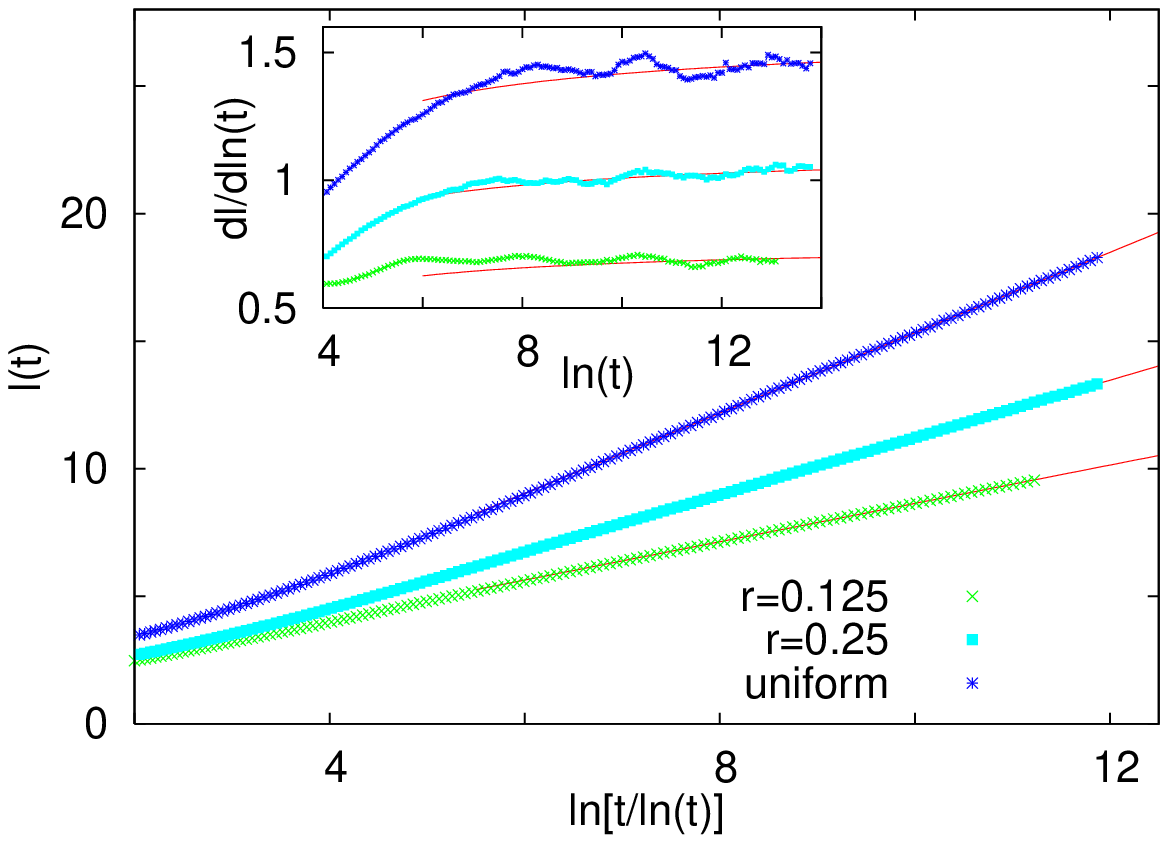}
\caption{\label{fig7} Left: The effective exponent $\beta_{\rm eff}(t)$ (open symbols)
  plotted against time for zero average force. Filled symbols
  show the effective exponents calculated from the variance of the
  current. The solid line corresponds to the form in Eq. (\ref{Jtc}).
Right: Time-dependence of the length scale $l(t)$. The effective exponents are
shown in the inset, where the solid lines corresponds to the form in Eq. (\ref{ltc}).}
\end{figure}

Next we turn to present numerical results obtained for zero average force 
($\mu=0$). 
The distributions of local currents for different times are shown in
Fig. \ref{fig5} and \ref{fig6}. The striking difference compared to the driven
phase is that, although the exponent $\beta$ is still finite ($\beta=1$) for
$\mu=0$, the distributions are broadening with increasing time.  
A rough scaling collapse can be achieved in terms of the scaling variable 
$\ln(J)/\ln(t)$. 
Earlier results on the distribution of the stationary
current in finite systems of size $L$ showed an approximate scaling collapse
for the scaling variable $\ln(J)/L^{1/2}$ \cite{jsi}. 
Taking into account the relation between time and length scale in
Eq. (\ref{xit}), this is consistent with our present results on the dynamical scaling. 
As can be seen in Fig. \ref{fig7}, the effective exponent $\beta_{\rm eff}(t)$  
overshoots the expected asymptotical value $1$ by a few percent. 
This is in accordance with that the scaling collapse of distributions is not
perfect and the shape of distributions is still slightly changing at the
numerically available time scales. We have also measured the variance of local
currents which enhances the contribution of large local currents compared to
the typical value defined in Eq. (\ref{typical}). These lie just in the still
deforming and thus poorly scaling part of distributions. The corresponding
effective exponents are approaching the theoretical value from below, see
Fig. \ref{fig7}. 
In the right panel of Fig. \ref{fig7}, the measured length scale is plotted
against time. As can be seen, the data are in good agreement with the law
given in Eq. (\ref{ltc}).

\section{Discussion}
\label{discussion}

The partially asymmetric simple exclusion process
with random-force disorder has been investigated in earlier studies
exclusively by Monte Carlo simulations and by a phenomenological random
 barrier description. 
In this work, a mean field approximation has been applied to this model and
the main focus was on the non-stationary phenomena. 
The mean field approximation leads to a system of deterministic, nonlinear
differential equations. This model is less complex than the original
stochastic process but still not tractable analytically. 
Nevertheless, it is appropriate for applying a 
phenomenological random barrier theory to it. 
According to our analytical mean field calculations, 
the key issue of the theory, namely the
current through a barrier has the same behavior as that has been assumed
intuitively for the original model in earlier works. 
This leads to the same large-scale stationary and
non-stationary behavior as has been conjectured for the original model. 
We have investigated the mean field model numerically, which is considerably
faster than performing Monte Carlo simulations and have found that the 
dynamics is satisfactorily described by the phenomenological theory. 
Since a good agreement between the phenomenology and results of Monte Carlo
simulations carried out on the original model has been found in earlier works, 
we conjecture that the mean field model belongs to the same universality class as
the original stochastic process does. This means that the static and dynamical
exponents in the driven phase, as well as the scaling relations for the case
of zero average force are identical. 
Our results show that in the presence of disorder the local
correlations are unimportant concerning the large scale behavior of the
system. 
This conclusion can be instructive for the investigation of other
transport processes with random-force disorder, where the simple mean
field approximation may give the correct large scale behavior.

\appendix
\section{}
Let us denote the stationary local density in the pure model by $\rho$ and
the deviation from the stationary density by $\epsilon_i(t)$, i.e.  
$\rho_i(t)=\epsilon_i(t)+\rho$. 
The spatially continuous limit of the evolution equations Eq. (\ref{evolution})
reads as 
\be 
\frac{\partial\epsilon}{\partial t}=D\frac{\partial^2\epsilon}{\partial
  x^2}-v\frac{\partial\epsilon}{\partial x} +
\lambda\epsilon\frac{\partial\epsilon}{\partial x},
\label{continuum}
\ee
where the constants $D$, $v$ and $\lambda$ are given in terms of the jump rates
as $D=(p+q)/2$, $v=(1-2\rho)(p-q)$ and $\lambda=2(p-q)$. 
The local current in the continuum limit takes the form 
\be
J(x,t)=J_{\infty}+v\epsilon-D\frac{\partial\epsilon}{\partial x}-\lambda\epsilon^2,
\label{current}
\ee
where $J_{\infty}=(p-q)\rho(1-\rho)$ is the current in the steady state. 
First, let us consider the simplest case $p=q$, which describes the symmetric simple
exclusion process. In this case, $v=\lambda=0$ and Eq. (\ref{continuum})
reduces to the diffusion equation. 
Consider a random initial density profile of the form
\be 
\epsilon(x,0)=\sum_{n=-\infty}^{\infty}s_n\delta(x-n),
\label{init}
\ee
where $\delta(x)$ is the Dirac delta distribution and the $s_n$ are 
independent binary random variables with the probability density
$f(s)=\frac{1}{2}\delta(s-1)+\frac{1}{2}\delta(s+1)$.
The solution of Eq. (\ref{continuum}) is then 
\be 
\epsilon(x,t)=\frac{1}{\sqrt{4\pi Dt}}
\sum_{n=-\infty}^{\infty}s_n\exp{\frac{-(x-n-vt)^2}{4Dt}}.
\label{diffsol}
\ee
The mean value of local quantities such as the deviation 
$\overline{\epsilon(t)}=\lim_{L\to\infty}\frac{1}{2L+1}\sum_{n=-L}^L\epsilon(n,t)$
at time $t$ can be calculated alternatively from
$\overline{\epsilon(t)}=\int\epsilon(0,t)f(s)ds$ since the $s_n$ are identically
distributed for all $n$.
For the mean deviation we obtain the obvious result
$\overline{\epsilon(t)}=0$, since $\overline{\epsilon(0)}=0$ and the
total mass $\int \rho(x,t)dx$ is conserved by Eq. (\ref{continuum}). 
The fluctuations of $\epsilon(x,t)$ are characterized by the variance, the
square of which can be easily calculated: 
\be
\overline{\epsilon^2(t)}=\frac{1}{4\pi Dt}\sum_{n=-\infty}^{\infty}\exp{\frac{-n^2}{2Dt}}\approx
\frac{1}{4\pi Dt}\int_{-\infty}^{\infty}\exp{\frac{-x^2}{2Dt}}dx=
(8\pi Dt)^{-1/2},
\ee
where the sum has been approximated by an integral. 
Thus, the typical deviation from the stationary density measured at a randomly
chosen site at time $t$ is in the order of $t^{-1/4}$. 
The typical local current can be estimated in a similar way. The mean value 
$\overline{\frac{\partial\epsilon}{\partial x}}$ is zero while the square of
  the variance is 
\beqn
\overline{\left(\frac{\partial\epsilon}{\partial x}\right)^2}=\frac{1}{4\pi Dt}\sum_{n=-\infty}^{\infty}\frac{n^2}{4D^2t^2}\exp{\frac{-n^2}{2Dt}}\approx
\nonumber
\\
\frac{1}{16\pi D^3t^3}\int_{-\infty}^{\infty}x^2\exp{\frac{-x^2}{2Dt}}dx=
\frac{1}{16}\sqrt{\frac{2}{\pi}}(Dt)^{-3/2}.
\eeqn
The typical current measured at a given bond at time $t$ is thus $O(t^{-3/4})$. 

In the case $p\neq q$, which corresponds to the asymmetric simple exclusion process, the Galilean transformation $x\to x+vt$ cancels the
second term on the r.h.s. of 
 Eq. (\ref{continuum}) and one obtains the noiseless
Burgers equation \cite{burgers}: 
\be
\frac{\partial\epsilon}{\partial t}=D\frac{\partial^2\epsilon}{\partial
  x^2}+\lambda\epsilon\frac{\partial\epsilon}{\partial x}.
\label{burgers}
\ee
This can be exactly solved by the Cole-Hopf transformation (see
e.g. \cite{stinchcombe1})
$w(x,t)=\exp{\frac{\lambda}{2D}\int^x\epsilon(x',t)dx'}$, which maps the
Burgers equation to the diffusion equation. 
The solution is obtained from 
\be
\epsilon(x,t)=\frac{2D}{\lambda}\frac{\partial\ln w}{\partial x}, 
\ee
with 
\be
w(x,t)=\int_{-\infty}^{\infty}\frac{1}{\sqrt{4\pi Dt}}\exp{\left(-\frac{(x-x')^2}{4Dt}+\frac{\lambda}{2D}\int^{x'}\epsilon(x'',0)dx''\right)}dx'.
\ee
Using the random initial condition given in Eq. (\ref{init}), the deviation of
the density at $x=0$ is given as 
\be 
\epsilon(0,t)= \frac{1}{Dt}
\frac{\int_{-\infty}^{\infty}x\exp{\left[-\frac{x^2}{4Dt}+\frac{\lambda}{2D}r(x)\right]}dx}
{\int_{-\infty}^{\infty}\exp{\left[-\frac{x^2}{4Dt}+\frac{\lambda}{2D}r(x)\right]}dx}, 
\label{expected}
\ee
where $r(x)=\int_0^x\epsilon(x',0)dx'$ is a piecewise constant function with
unit jumps at integers whereas for non-integers it is given by 
$r(x)=-\frac{s_0}{2}+\sum_{n=0}^{[x]}s_n$, where $[x]$ denotes the integer
part of $x$. Thus $r(x)$ can be regarded as a random walk which
makes jumps at integer ``times'' $x$.
It is easy to see that the mean value of the deviation
$\overline{\epsilon(t)}$ is zero since
$\overline{\exp{\frac{\lambda}{2D}r(x)}}$ is an even function of $x$ due to 
 ${\rm Prob}[r(x)]={\rm Prob}[r(-x)]$.
The typical value of the deviation at time $t$ can be obtained as
follows. First notice that the r.h.s. of  Eq. (\ref{expected}) can be
regarded as the expected value of $x$ which has the (unnormalized)  
weight function $\exp{F(x)}$ with
\be
F(x)=-\frac{x^2}{4Dt}+\frac{\lambda}{2D}r(x).
\label{F}
\ee 
The dominant contribution to this expected value comes from the interval where
$F(x)$ is maximal, since otherwise the weight $\exp{F(x)}$ is negligible.
So we have the approximate relation 
$\epsilon(0,t)\sim \frac{1}{Dt}x_{\rm max}$, where $x_{\rm max}$ is the location of the
maximum of $F(x)$. (In case there are many maxima, $x_{\rm max}$ is their mean
value.)
The distribution of $x_{\rm max}$ is symmetric around zero and the dependence
of its magnitude on time can be obtained by taking into account that the
variance of $r(x)$ which characterizes its typical magnitude is 
$\sqrt{\overline{r^2(x)}}=\sqrt{[x]+1/4}$, i.e. proportional to $\sqrt{x}$ for
large $x$. Replacing $r(x)$ in Eq. (\ref{F}) by $\sqrt{x}$,  we obtain a non-random function the maximum of which
is at $(2\lambda t)^{2/3}$. Thus the width of the distribution of 
$x_{\rm max}$ is in the order of $t^{2/3}$ and we obtain
finally that the typical deviation from the stationary density measured at a
given site scales with time as $\epsilon_{\rm typ}(t)\sim t^{-1/3}$.    
By a similar calculation one can show that the expected value of
$\frac{\partial\epsilon}{\partial x}$ is of the order of $t^{2/3}$.
Now we can turn to the analysis of the fluctuations of the current. 
If $\rho\neq 1/2$ then $v\neq 0$ and the fluctuations are dominated by the
term $v\epsilon$, see Eq. (\ref{current}). Thus the magnitude of the typical
local current (relative to the stationary current) scales as
$(J-J_{\infty})_{\rm typ}(t)\sim t^{-1/3}$. 
If, however, $\rho=1/2$, the above term is zero and the fluctuations are
determined by the other two terms leading to $(J-J_{\infty})_{\rm typ}(t)\sim t^{-2/3}$.

\ack
The author thanks useful discussions with F. Igl\'oi, I. Kov\'acs and
G. \'Odor. 
This paper was supported by the J\'anos Bolyai Research Scholarship of the
Hungarian Academy of Sciences and by the Hungarian National Research Fund
under grant no. OTKA K75324.



\section*{References}
\begin{thebibliography}{99}


\bibitem{motors}
J. Howard, {\it Mechanics of Motor Proteins and the Cytoskeleton} 
(Sinauer, Sunderland, 2001)

\nonum
Schliwa M and Woehlke G, 2003 {\it Nature} {\bf 422} 759 

\nonum
Gross S P, 2004 {\it Phys. Biol.} {\bf 1} R1 

\nonum
Chowdhury D, Schadschneider A, Nishinari K
2005 {\it Physics of Life Reviews} (Elsevier, New York) vol. 2, p. 318 


\bibitem{mcdonald} MacDonald C T, Gibbs J H and Pipkin A C, 1968 
{\it Biopolymers} {\bf 6} 1

\bibitem{spitzer} 
Spitzer F, 1970 {\it Adv. Math.} {\bf 5} 246 


\bibitem{liggett} Liggett T M 1999 {\it Stochastic interacting systems:
  contact, voter, and exclusion processes} (Berlin, Springer)

\bibitem{zia}
Schmittmann B and Zia R K P 1995 in 
{\it Phase Transitions and Critical Phenomena}, 
vol. 17, edited by Domb C and Lebowitz J L (Academic, London)

\bibitem{schutzreview} Sch\"utz G M 2001 in {\it Phase  Transitions and
    Critical Phenomena}, vol. 19, edited by Domb C and Lebowitz J L (Academic, San Diego)

\bibitem{be}
Blythe R A, Evans M R 2007 {\it J. Phys. A Math. Theor.} {\bf 40} R333

\bibitem{ramaswamy} 
Ramaswamy R, Barma M 1987 {\it J. Phys. A: Math. Gen.} {\bf 20} 2973

\bibitem{stanley}
Koscielny-Bunde E, Bunde A, Havlin S, and Stanley H E,
1988 {\it Phys. Rev. A} {\bf 37} 1821

\bibitem{tripathy} Tripathy G, Barma M 1997 {\it Phys. Rev. Lett.} {\bf 78}
  3039; 1998 {\it Phys. Rev. E} {\bf 58} 1911

\bibitem{goldstein}
Goldstein S, Speer E R 1998 {\it Phys. Rev. E} {\bf 58} 4226

\bibitem{kolwankar}
Kolwankar K M, Punnoose A 2000 {\it Phys. Rev. E} {\bf 61} 2453

\bibitem{krug2}
Krug J, 2000 {\it Braz. J. Phys.} {\bf 30} 97

\bibitem{hs}
Harris R J, Stinchcombe R B 2004 {\it Phys. Rev. E} {\bf 70} 016108

\bibitem{jsi}
Juh\'asz R, Santen L and Igl\'oi F, 2005 {\it Phys. Rev. Lett.} {\bf 94} 010601;
2006 {\it Phys. Rev. E} {\bf 74} 061101

\bibitem{jli06} Juh\'asz R, Lin Y-C, Igl\'oi F 2006 {\it Phys. Rev. B} {\bf 73} 224206

\bibitem{barma} 
Barma M, 2006 {\it Physica A} {\bf 372} 22

\bibitem{schadschneider}
Greulich P, Schadschneider A 2008 {\it J. Stat. Mech.} P04009

\bibitem{burgers}
Burgers J M 1974 {\it The Nonlinear Diffusion Equation} (Riedel, Boston)

\bibitem{blythe} 
Blythe R A, Evans M R, Colaiori F, Essler F H L 2000
{\it J. Phys. A: Math. Gen.} {\bf 33} 2313

\bibitem{ir}
Igl\'oi F, Rieger H 1998 {\it Phys. Rev. E} {\bf 58} 4238


\bibitem{bouchaud} For a review see: Bouchaud J P, Georges A 1990 {\it Phys. Rep.} {\bf 195} 127


\bibitem{numrec}
Press W H, Teukolsky S A, Wetterling W T, Flannery B P 1992 {\it Numerical Recipes in C} (Cambridge University Press, Cambridge)


\bibitem{stinchcombe1}
Stinchcombe R B 2001 {\it Adv. in Phys.} {\bf 50} 431 


\end{thebibliography}
\end{document}